\documentclass[psfig]{article}
\usepackage{amssymb}

\newcommand{\be}{\begin{equation}}
\newcommand{\ee}{\end{equation}}
\newcommand{\bea}{\begin{eqnarray}}
\newcommand{\eea}{\end{eqnarray}}

\title{Abelian Chern-Simons Vortices and Holomorphic Burgers' Hierarchy}

\author{Oktay K. PASHAEV  and
Zeynep Nilhan GURKAN \\ \\
Department of Mathematics,
Izmir Institute of Technology \\
Urla-Izmir, 35430, Turkey}

\begin{document}
\maketitle

\begin{abstract}

The Abelian Chern-Simons Gauge Field Theory in 2+1 dimensions and
its relation with holomorphic Burgers' Hierarchy is considered. It
is shown that the relation between complex potential and the
complex gauge field as in incompressible and irrotational
hydrodynamics, has meaning of the analytic Cole-Hopf
transformation, linearizing the Burgers Hierarchy in terms of the
holomorphic Schr\"odinger Hierarchy. Then the motion of planar
vortices in Chern-Simons theory, appearing as pole singularities
of the gauge field, corresponds to motion of zeroes of  the
hierarchy. Using boost transformations of the complex Galilean
group of the hierarchy, a rich set of exact solutions, describing
integrable dynamics of planar vortices and vortex lattices in
terms of the generalized Kampe de Feriet and Hermite polynomials
is constructed. The results are applied to the holomorphic
reduction of the Ishimori model and the corresponding hierarchy,
describing dynamics of magnetic vortices and corresponding
lattices in terms of complexified Calogero-Moser models.
Corrections on two vortex dynamics from the Moyal space-time
non-commutativity in terms of Airy functions are found.

\end{abstract}

~% THEOREM Environments ---------------------------------------------------
\newtheorem{thm}{Theorem}[subsection]
\newtheorem{cor}[thm]{Corollary}
\newtheorem{lem}[thm]{Lemma}
\newtheorem{prop}[thm]{Proposition}
\newtheorem{defn}[thm]{Definition}
\newtheorem{rem}[thm]{Remark}
\newtheorem{prf}[thm]{Proof}

\section{Classical Ferromagnetic Model in Continuous Media}

In order to solve the so-called momentum problem in planar
ferromagnets a model of delocalized electrons has been introduced
by Volovik, where the restoration of the linear momentum density
has appeared by using hydrodynamical variables, the density and
the normal velocity of the fermionic liquid. On this basis a
simple model of ferromagnetic fluid or a spin-liquid, modifying
the phenomenological Landau-Lifshitz equation, has been proposed
in \cite{MPS1}. In the model, the magnetic variable is described
by the classical spin vector $\vec{S} = \vec{S}(x, y, t)$ valued
on two dimensional sphere: $\vec{S}^2 = 1$, and the hydrodynamic
variable is velocity $\vec{v}(x, y, t)$ of the incompressible
flow.  For particular anisotropic case of the space metric we have
the system
\begin{equation}
\vec{S}_{t}+\upsilon_{1}\partial_{1}\vec{S}-\upsilon_{2}\partial_{2}\vec{S}=
\vec{S}\times(\partial_{1}^{2}-\partial_{2}^{2})\vec{S}\label{maintopmag1}
\end{equation}
\begin{equation}
\partial_{1}\upsilon_{2}-\partial_{2}\upsilon_{1}=
2\vec{S}(\partial_{1}\vec{S}\times\partial_{2}\vec{S})\label{maintopmag2}
\end{equation}
The first equation is the Heisenberg model where time derivative
$\partial/\partial t$ is replaced by the material derivative $
\frac{D}{Dt} = \frac{\partial}{\partial t} + (\vec{v}\nabla)$,
while the second one is relation between hydrodynamic and spin
variables called the Mermin-Ho relation \cite{Mermin}. It relates
vorticity of the flow with topological charge density, or the
winding number for spin. In this case the next theorem is valid
\cite{MPS3}.

{\bf Theorem:} For the flow constrained by the incompressibility
condition
\begin{equation}
\partial_{1}\upsilon_{1}+\partial_{2}\upsilon_{2}= 0 ,\label{incompres}
\end{equation}
the conservation law, $
\partial_{t}J_{0}+\partial_{2}J_{2}-\partial_{1}J_{1}=0$,
 holds, where
 \begin{eqnarray}
 J_{0}&=&(\partial_{1}\vec{
S})^2+(\partial_{2}\vec{S})^2,\label{energy}
\\
J_{1}&=&-2\partial_{1}\vec{S}\cdot\vec{S} \times
(\partial_{1}^{2}-\partial_{2}^{2})\vec{S} +
v_{1}J_{0}+2\vec{S}\cdot(\partial_{1}\vec{S} \times
\partial_{2}^{2}\vec{S}
-\partial_{1}\partial_{2}\vec{S}\times\partial_{2}
\vec{S})\nonumber \\
J_{2}&=& 2\partial_{2}\vec{S}\cdot\vec{S} \times
(\partial_{1}^{2}-\partial_{2}^{2})\vec{S} + v_{2}J_{0} -2
\vec{S}\cdot(\partial_{1}^{2}\vec{S} \times
\partial_{1}\partial_{2}\vec{S} -\partial_{1}\vec{S}
\times\partial_{2}\vec{S}).\nonumber
\end{eqnarray}
Due to the above theorem, for the incompressible flow
(\ref{incompres}) the functional (\ref{energy}) (the "energy"
functional) $E = \int\int J_0 d^2 x$  is conserved quantity,
bounded $E \geq 8\pi|Q|$ (Bogomolnyi Inequality) by the
topological charge $Q$ of a spin configuration. This inequality is
saturated for spin configurations satisfying the first order
system, the Belavin-Polyakov self-duality equations
\begin{equation}
\partial_{i}\vec{S}\pm\epsilon_{ij}\vec{S}\times
\partial_{j}\vec{S}= 0\label{selfdual}
\end{equation}
In contrast to the static case our inequalities are valid for time
dependent fields.  The stereographic projections of the spin phase
space are given by formulas
\begin{equation}
S_+=S_{1}+ iS_{2}=\frac{2\zeta}{1+|\zeta|^{2}} ,\,\,\,\,\,\,\
S_{3}=\frac{1-|\zeta|^{2}}{1+|\zeta|^{2}}\label{spanal3}
\end{equation} where
$\zeta$ is complex valued function. By complex derivatives
$
\partial_{z}=\frac{1}{2}(\partial_{1}-i\partial_{2})$, $\partial_{\bar{z}}=\frac{1}{2}(\partial_{1}+i\partial_{2})
$, the self-duality equations (\ref{selfdual}) in stereographic
projection form become
 the analyticity  $\zeta_{\bar{z}}(z, t) =0$ or
  the anti-analyticity
  $\zeta_{z}(\bar z, t)=0$ conditions.
It is easy to show by direct calculations that for incompressible
flow (\ref{incompres}) the holomorphic constraint $\zeta_{\bar
z}(z,t)=0$ is compatible with the time evolution $
\frac{\partial}{\partial t}\zeta_{z}=0$.

\section{Holomorphic Reduction of Ishimori Model}
From the theorem and the proposition presented above we have seen
that incompressible flow admits existence of positive energy
functional minimized by holomorphic reduction and this reduction
is compatible with the time evolution. It suggests to solve the
incompressibility conditions explicitly. So we consider the
topological magnet model (\ref{maintopmag1}),(\ref{maintopmag2})
with incompressibility condition solved in terms of real function
$\psi$, the stream function of the flow,
 $v_{1}=\partial_{2}\psi$, $v_{2}=-\partial_{1}\psi$. Then we have
 analytic reduction of
 the Ishimori model \cite{Ishimori}
\begin{equation}
i\zeta_{t}-2\psi_{{z}}\zeta_{{z}}
+2\zeta_{{z}{z}}-4\frac{{\zeta}}{1+|\zeta|^{2}}\zeta_{{z}}^{2}=0\label{holomIshimori1},
\end{equation}
\begin{equation}
\psi_{z\bar{z}}=-2\frac{\bar{\zeta}_{\bar
z}\zeta_{{z}}}{(1+|\zeta|^{2})^{2}} \label{holomIshimori2
}.\end{equation} Choosing the stream function \be \psi=2\ln
(1+|\zeta|^2) \ee equation (\ref{holomIshimori2 }) is satisfied
automatically, while from (\ref{holomIshimori1}) we have the
holomorphic Schr\"odinger equation \be i\zeta_t+2\zeta_{zz}=0
\label{hSchrodinger}.\ee Every zero of function $\zeta$ in complex
plane $z$ determines magnetic vortex of the Ishimori model. The
spin vector at  center of the vortex is $\vec S = (0,0,1)$ while
at infinity $\vec S = (0,0,-1)$. Then a motion of zeroes of
equation (\ref{hSchrodinger}) determines the motion of magnetic
vortices in the plane. From another side, if we consider analytic
function \be f(z,t) = \frac{\Gamma}{2\pi i}Log\, \zeta
(z,t)\label{compot}\ee as the complex potential of an effective
flow \cite{Lav}, then every zero of function $\zeta$ corresponds
to hydrodynamical vortex of the flow with intensity $\Gamma$, and
to the simple pole singularity of complex velocity \be u (\bar z,
t)= \bar f_{\bar z} = \frac{i\Gamma}{2\pi}(Log \,\bar \zeta)_{\bar
z}. \ee But the last relation has meaning of the holomorphic
Cole-Hopf transformation, according to which the complex velocity
is subject to the holomorphic Burgers' equation \be iu_t+
\frac{8\pi i}{\Gamma} uu_{\bar z}= 2u_{\bar z \bar z}
\label{aholBurgers} \ee Thus, every magnetic vortex of the
Ishimori model corresponds to hydrodynamical vortex of the
anti-holomorphic Burgers' equation. Moreover, relation
(\ref{compot}) written in the form \be \zeta = e^{\frac{2\pi
i}{\Gamma} f} = e^{\frac{2\pi i}{\Gamma} (\phi + i \chi)} =
\sqrt{\rho}\, e^{\frac{2\pi i}{\Gamma} \phi}\ee shows that the
effective flow is just the Madelung representation for the linear
holomorphic Schr\"odinger equation (\ref{hSchrodinger}), where
functions $\phi$ and $\chi$ are the velocity potential and the
stream function correspondingly.

\section{N-Vortex System}
The system of N magnetic vortices is determined by  N simple
zeroes
\begin{equation}
 \zeta({z},t)=\prod_{k=1}^{N}({z}-{z}_{k}(t))
 \end{equation}
positions of which according to (\ref{hSchrodinger})  are subject
to the system
\begin{equation}
\frac{d {z}_{k}}{dt}=\sum_{k(\neq
l)=1}^N\frac{4i}{{z}_{k}-{z}_{l}}\label{ishvortex}.
\end{equation}
In one space dimension this system has been considered first in
\cite{Chood}, (see also \cite{Calogero}) for moving poles of
Burgers' equation, determined by zeroes of the heat equation.
However, complexification of the problem has several advantages.
First of all the root problem of algebraic equation degree N is
complete in the complex domain as well as the moving singularities
analysis of differential equations. In contrast to one dimension,
in this case the pole dynamics in the plane becomes time
reversible (see below) and has interpretation of the vortex
dynamics. Moreover generalization (non-integrable) of the system
(\ref{ishvortex}) to the case of three particles with different
strength has been studied in \cite{CalSanSom} to explain the
transition from regular to irregular motion as travel on the
Riemann surfaces.

 Below in Section 7 we show that solution of this system is determined by
 N complex constants of motion, this is why the dynamics of vortices in Ishimori
model is integrable. In fact the system (\ref{ishvortex}) admits
mapping to the complexified Calogero-Moser (type I) N particle
problem \cite{Perelomov}, \cite{Calogerob}. For these we
differentiate once and use the system again to have Newton's
equations
\begin{equation}
\frac{d^{2}}{dt^{2}}{z}_{k}= \sum_{l=1(\neq
k)}\frac{32}{({z}_{j}-{z}_{k})^{3}}\label{CalMoser}.
\end{equation}
These equations have Hamiltonian form
 with Hamiltonian function
\begin{equation}
H=\frac{1}{2}\sum_{k=1}^{n}p^{2}_{k}+\sum_{k<l}\frac{16}{(z_{k}-z_{l})^2}
\ee  and  admit the Lax representation, from which follows the
hierarchy of constants of motion in involution $ I_k = tr
L^{k+1}$. Recently complexification of the classical
Calogero-Moser model and holomorphic Hopf equation has been
considered in connection with limit of an infinite number of
particles, leading to quantum hydrodynamics and quantum
Benjamin-Ono equation \cite{Wiegmann}. From another side
holomorphic version of the Burgers equation is considered in
\cite{Bonami} to prove existence and uniqueness of the non-linear
diffusion process for the system of Brownian particles with
electrostatic repulsion when the number of particles increases to
infinity.

\section{ Integrable N-particle Problem for N-Vortex Lattices}

The function $\zeta$ of the form \be \zeta (z,t) = \sin (z - z_k
(t)) = (z - z_k (t)) \prod_{n = 1}^\infty \left( 1 - \frac{(z -
z_k (t))^2}{n^2 \pi^2}\right)\ee has periodic infinite set of
zeroes and determines the vortex lattice. First for
(\ref{hSchrodinger}) we consider the system of N vortex chain
lattices periodic in $x$ \be \zeta(z,t)= e^{-2iN^2t} \prod_{k=1}^N
\sin (z-z_k(t)) \ee so that positions are subject to the first
order system \be \dot{z_k}=2 i\sum_{l=1(\neq k)} \cot(z_k -
z_l).\label{n lattices}\ee Differentiating this system once  in
time we get the second order equations of motion in the Newton's
form \be \ddot{z}_k = 32\sum_{l} \frac{\cot
({z}_k-{z}_l)}{\sin^2({z}_k-{z}_l)} \ee with the Hamiltonian
function of the Calogero-Moser type II model \cite{Perelomov}
\begin{equation}
H=\frac{1}{2}\sum_{k=1}^N{p}_k ^2 +
\sum_{k<l}\frac{16}{\sin^2({z}_k-{z}_l)} .\end{equation} For
periodic in $y$ lattices \be \zeta(z,t)= e^{2iN^2t} \prod_{k=1}^N
\sinh (z-z_k(t)) \ee we get the Calogero-Moser type III Model
\begin{equation}
H=\frac{1}{2}\sum_{k=1}^N{p}_k ^2 +
\sum_{k<l}\frac{16}{\sinh^2({z}_k-{z}_l)}.
\end{equation}

\section{Complex Galilean Group and Vortex Generations}
Complex Galilean Group is generated by algebra \be
[P_0,P_z]=0,\,\,\,\, [P_0,K]=4iP_z ,\,\,\,\, [P_z,K]=-i \ee where
the energy and momentum operators are $ P_0=-i \partial_t$,
$P_z=-i\partial_z $ correspondingly, while the Galilean Boost is
operator\be K=z+4it
\partial_z.\ee
The Schr\"odinger operator \be S=i\partial_t+2\partial^2_z \ee
corresponds to the dispersion relation $ P_0=-2P^2_z $ and is
commuting with Galilean group \be [P_0,S]=0,\,\,\,\, [P_z,S]=0
,\,\,\,\, [K,S]=0. \ee From the theory of dynamical symmetry, if
exists operator $W$ such that \be [S,W]=0 \Rightarrow
S(W\Phi)=W(S\Phi)=0 \ee then it transforms solution $\Phi$ of the
Schr\"odinger equation to another solution $W\Phi$. It shows that
Galilean generators provide dynamical symmetry for the equation.
Two of them are evident: time translation $ P_0:e^{it_0P_0}
\Phi(z,t)=\Phi(z,t+t_0)$ and the complex space translation  $
P_z:e^{it_0P_z} \Phi(z,t)=\Phi(z+z_0,t)$. While the Galilean boost
creates new zero (new vortex in C) \be
\Psi(z,t)=K\Phi(z,t)=(z+4it\partial_z)\Phi(z,t)\label{boost}.\ee
Starting from evident solution $\Phi=1$ we have the chain of
n-vortex solutions, $K\cdot 1= z = H_1(z,2it)$, $ K^2 \cdot 1=
z^2+4it=H_2(z,2it)$, $K^3\cdot 1= z^3+12it=H_3(z,2it)$,...,$
K^n\cdot 1= H_n(z,2it)$ in terms of the Kampe de Feriet
Polynomials \cite{Dattoli-97} \be H_n(z,it)=n!
\sum_{k=0}^{[n/2]}\frac{(it)^k z^{n-2k}}{k! (n-2k)!}.\ee They
satisfy the recursion relations \be H_{n+1}(z,it)= \left(z+2it
\frac{\partial}{\partial z}\right)H_n(z,it), \ee \be
\frac{\partial}{\partial z} H_n(z,it)=n H_{n-1}(z,it) \ee and can
be written in terms of Hermite polynomials \be
H_n(z,2it)=(-2it)^{n/2}H_n\left(\frac{z}{2\sqrt{-2it}}\right)\ee

Let $w_n^{(k)}$ is the k-th zero of the Hermite polynomial,
$H_n(w_n^{(k)})=0$  , then  evolution of corresponding vortex is
given by \be z_k(t)=2w_n^{(k)}\sqrt{-2it}.\label{2vortex}\ee Under
the time reflection $t \rightarrow -t$ position of the vortex
rotates on 90 degrees $z_k \rightarrow z_k e^{i\pi/2}$. The last
one is also symmetry of the vortex equations (\ref{ishvortex}).
Using formula
 \be H_n(z,2it)=\exp
\left(it\frac{\partial^2}{\partial z^2}\right)z^n\ee and
superposition principle we have solution $$
\Phi(z,t)=\sum_{n=0}^{\infty}a_n
H_{n}(z,2it)=\sum_{n=0}^{\infty}a_n \exp
\left(2it\frac{\partial^2}{\partial z^2}\right)z^n=\exp
\left(2it\frac{\partial^2}{\partial
z^2}\right)\sum_{n=0}^{\infty}a_n z^n $$ So if $\chi (z) =
\sum_{n=0}^{\infty}a_n z^n$ is any analytic function, then $
\Phi(z,t)=\exp \left(2it\frac{\partial^2}{\partial
z^2}\right)\chi(z)$ is solution, determined by integrals of motion
$a_0, a_1,...$. Therefore for polynomial degree $n$ describing
evolution of $n$ vortices, we have $n$ complex integrals of
motion.

The generating function of the Kampe de Feriet Polynomials \be
\sum_{n=0}^{\infty}\frac{k^n}{n!}H_{n}(z,it)=e^{kz+ik^2t}
\label{genfuncKF}\ee is also solution of the plane wave type. If
we exponentiate the Galilean boost $ e^{i\lambda K}= e^{i\lambda
(z+4it\partial_z)}$ and factorize it  by Baker-Hausdorf formula
$e^{A+B}=e^B e^A e^{\frac{1}{2}[A,B]}$, so that $ e^{i\lambda
K}=e^{i\lambda z+2i\lambda^2 t} e^{-4\lambda t
\partial_z}$, then
applied on a solution  $\Phi(z,t)$ it gives \be e^{i\lambda
K}\Phi(z,t)=e^{i\lambda z+ 2i\lambda^2 t}\Phi(z-4\lambda t,t)\ee
the Galilean boost with velocity $4\lambda$, where the generating
function of vortices (\ref{genfuncKF}) appears as the 1-cocycle.

The Galilean boost (\ref{boost}) $ \Psi(z,t)=(z+4it
\partial_z)\Phi $ connecting two solutions of the holomorphic Schr\"odinger
equation (\ref{hSchrodinger}) generates the auto-B\"acklund
transformation : \be v=u+ \frac{i\Gamma}{2\pi}\partial_{\bar z}
\ln ({\bar z} - \frac{8\pi t}{\Gamma} u)\ee between two solutions
 \be u (\bar z, t)=\frac{i\Gamma}{2 \pi}\frac{\bar\Phi_{\bar z}}{\bar\Phi} ,\,\,\,\,\,
v (\bar z, t)=\frac{i\Gamma}{2 \pi}\frac{\bar\Psi_{\bar
z}}{\bar\Psi}\ee
 of the anti-holomorphic Burgers equation
(\ref{aholBurgers}).

As an example we consider double lattice solution \be
\zeta(z,t)=e^{-8it}\sin (z-z_1(t))\sin(z+z_1(t)) \ee where $ \cos
2z_1=re^{8it}$, $r$ is a constant. Applying the boost
transformation (\ref{boost}) we have solution describing collision
of a vortex with the double lattice \be
\Psi(z,t)=\left(z+4it\frac{\partial}{\partial z}\right)\zeta(z,t)
\ee Generalizing we have N-vortices interacting with M-vortex
lattices \be \Psi(z,t)=e^{iMt}\left(z+4it\frac{\partial}{\partial
z}\right)^N \prod_{k=1}^M \sin(z-z_k(t))\ee where $z_1,...,z_k$
are subject
 to the system (\ref{n lattices}).

\section{Abelian Chern-Simons Theory and Complex Burgers' Hierarchy}
Now we show as the anti-holomorphic Burgers hierarchy appears in
the Chern-Simons gauge field theory. The Chern-Simons functional
is defined as follows\be S(A)=\frac{\kappa}{4\pi}\int_M A \wedge
dA= \frac{\kappa}{4\pi}\int \varepsilon^{\mu \nu
\lambda}A_{\mu}F_{\nu \lambda}\ee where $M$ is an oriented
three-dimensional manifold, $A$ is U(1) gauge connection, $\kappa$
the coupling constant - the statistical parameter. In the
canonical approach $M=\Sigma_2 \times R$, where $R$ we interpret
as a time. Then  $A_{\mu}=(A_0,A_i)$, $(i=1,2)$, where $A_0$ is
the time component and the action takes the form \be
S=-\frac{\kappa}{4\pi}\int dt \int_{\Sigma}\epsilon^{ij}\left(A_i
\frac{d}{dt}A_j-A_0 F_{ij}\right)\ee In the first order formalism
it implies that the Poisson brackets \be
\{A_i(x),A_j(y)\}=\frac{4\pi}{\kappa}\epsilon_{ij}\delta(x-y) \ee
and the Hamiltonian $H = A_0 \epsilon^{ij}F_{ij}$. The last one is
weakly vanishing $H \approx 0$ due to the Chern-Simons Gauss law
constraint \be
\partial_1A_2-\partial_2A_1=0 \Leftrightarrow F_{ij}=0\label{Gauss}
\ee Then the evolution is determined by Lagrange multipliers
$A_0$: $
\partial_0 A_1=\partial_1 A_0$,
$\partial_0 A_2=\partial_2 A_0$. Due to the gauge invariance
$A_{\mu}\rightarrow A_{\mu}+\partial_{\mu}\lambda$, to fix the
gauge freedom we choose the  Coulomb gauge condition: $div \vec A
=0$. In addition we have the Chern-Simons Gauss law (\ref{Gauss}):
$rot \vec A =0$ . These two equations are identical to the
incompressible and irrotational hydrodynamics. Solving the first
equation  in terms of the velocity potential $\varphi$:
$A_k=\partial_k\varphi$, $(k=1,2)$, and the second one  in terms
of the stream function $\psi$: $A_1=\partial_2\psi$ and
$A_2=-\partial_1\psi$; we have the Cauchy- Riemann Equations:
$\partial_1 \varphi=\partial_2\psi$,
$\partial_2\varphi=-\partial_1\psi$. So these two functions are
harmonically conjugate and the complex potential
$f(z)=\varphi(x,y)+ i\psi(x,y)$ is analytic  function of $z = x+
iy$: $\partial f / \partial \bar z =0$. Corresponding "the complex
gauge potential" $A=A_1+iA_2=\overline{f'(z)}$ is an anti-analytic
function. In analogy with hydrodynamics,  the logarithmic
singularities of the complex potential\be f(z,t)=\frac{1}{2\pi
i}\sum_{k=1}^N \Gamma_k Log (z-z_k(t))\ee determine poles of the
complex gauge field \be A=\frac{i}{2\pi}\sum_{k=1}^N
\frac{\Gamma_k}{\bar z - \bar z_k(t)}\ee describing point vortices
in the plane. Then the corresponding "statistical" magnetic field
\be B=\partial_1A_2-\partial_2A_1=-\Delta\psi=-\Delta \Im f(z) \ee
where $\Delta$ is the Laplacian,  determined by the stream
function
 \be \psi=-\frac{1}{2\pi}\sum_{k=1}^N
\Gamma_k Log |z-z_k(t)|\ee is \be B=\frac{1}{2\pi}\sum_{k=1}^N
\Gamma_k \Delta Log |z-z_k(t)| =\sum_{k=1}^N \Gamma_k \delta(\vec
r-\vec r_k(t)) \label{B delta}.\ee Corresponding total magnetic
flux is\be \int_{R^2} \int B d^2 x=\sum_{k=1}^N \int\int\Gamma_k
\delta(\vec r-\vec r_k(t))d^2x=\Gamma_1+\Gamma_2+...+\Gamma_N.\ee

The above relation (\ref{B delta}) has interpretation as  the
Chern-Simons Gauss law \be B=\frac{1}{\kappa}\bar \psi
\psi=\frac{1}{\kappa}\rho \ee for point particles located at $\vec
r_k(t)$ with density \be \rho= \sum_{k=1}^N \Gamma_k \delta(\vec
r-\vec r_k(t)) \ee (with masses $\Gamma_1,\Gamma_2,...,\Gamma_N$).
Then magnetic fluxes are superimposed on particles and has meaning
of anyons. As a result, an integrable evolution of the complex
gauge field singularities (vortices) would lead to the integrable
evolution of anyons. Evolution of the anti-holomorphic complex
gauge potential is determined by equation, $\partial_0 A=
2\partial_{\bar z}A_0$,  where as follows function $A_0$ is
harmonic $\Delta A_0 = 0$, and is given by $
A_0=\frac{1}{2}[F_0(\bar z,t)+\bar F_0(z,t)]$. Then the evolution
equation is \be
\partial_0A=\partial_{\bar z} F_0. \ee
Let \be F_0=\sum_{n=0}^{\infty}c_n F_0^{(n)}(\bar z,t)\ee where
\be F_0^{(n)}(\bar z,t)=(\partial_{\bar z}+A(\bar z,t))^n\cdot
1\ee then for arbitrary positive integer $n$ we have the
anti-holomorphic Burgers Hierarchy \be
\partial_{t_n}A(\bar z,t)=\partial_{\bar z}[(\partial_{\bar
z}+A(\bar z,t))^n\cdot 1]\label{n-Burgers}.\ee Using the recursion
operator $R = \partial_{\bar z} + \partial_{\bar z}
A\partial_{\bar z}^{-1} $ we write it in the form \be
\partial_{t_n}A = R^{n-1}\partial_{\bar z} A.\ee

The above hierarchy can be linearized by anti-holomorphic
Cole-Hopf transformation for the complex gauge field \be
A=\frac{\bar \Phi_{\bar z}}{\bar{\Phi}}=(\ln \bar{\Phi})_{\bar
z}=\overline{(f(z,t))}_{\bar z}\label{compgaugefield}\ee in terms
of the holomorphic Schr\"odinger(Heat) Hierarchy \be
\partial_{t_n}\Phi=\partial_z^n \Phi \label{holSchrodingerHierarch}.\ee
For $n=2$ the second member of the hierarchy is just
(\ref{hSchrodinger}) and zeroes of this equation corresponds to
magnetic vortices of the Ishimori model. The relation between
$\Phi$ and complex potential $f$ has meaning of the Madelung
representation for the hierarchy\be
\Phi(z,t)=e^{f(z,t)}=e^{\varphi+i\psi}=(e^\varphi)e^{i\psi}=\sqrt{\rho}
e^{i\psi}. \ee Therefore hierarchy of equations  for $f$ is the
Madelung form of the holomorphic Schr\"odinger hierarchy \be
\partial_{t_n} f=(\partial_z+\partial_z f)^n\cdot 1= e^{-f}
\partial_z^n e^f\ee
or \be
\partial_{t_n}(e^f)=\partial_z^n(e^f)\ee
which is the potential Burgers' hierarchy. We have the next Linear
Problem for the Burgers hierarchy \be \Phi_z=\bar A
\Phi,\,\,\,\,\,\,\,\, \Phi_{t_n}= \partial_z^n \Phi .\ee It can be
written as the Abelian zero-curvature representation for the
holomorphic Burgers hierarchy, $\partial_{t_n}U -
\partial_{\bar z} V_n = 0$,
where $U = A$, $V_n = (\partial_{\bar z} + A)^n \cdot 1$. For the
$N$-vortices of equal strength \be \Phi(z,t)=e^f= \prod_{k=1}^N
(z-z_k(t))\ee positions of the vortices correspond to zeroes of
$\Phi(z,t)$. As a result the vortex dynamics, leading to
integrable anyon dynamics, is related to motion of zeroes subject
to the vortex equations (\ref{ishvortex}) for $n=2$ case and for
arbitrary $n$ to equation \be -\frac{dz_k (t_n)}{dt_n} =
Res_{z=z_k}\left(\partial_z + \sum_{l=1}^N
\frac{1}{z-z_l(t_n)}\right)^n \cdot 1,\,\,\,(k =1,...,N).\ee

\section{Galilean Group Hierarchy and Vortex Solutions}
Now we consider complex Galilean Group hierarchy\be
[P_0,P_z]=0,\,\,\,\,\,[P_0,K_n]=i^n n P_z^{n-1},\,\,\,\,\,
[P_z,K_n]=-i \ee where hierarchy of the boost transformations is
generated by
 \be K_n=z+nt\partial_z^{n-1}\ee
is commuting with the holomorphic $n$-Schr\"odinger equation\be
S_n=\partial_t-\partial_z^n.\ee As a result, application of $K_n$
to solution $\Phi$ creates solution with additional vortex \be
\Psi(z,t)=K_n \Phi(z,t)=(z + n t
\partial_z^{n-1})\Phi(z,t)\label{n-boost}.\ee
For particular values we have $ K_n\cdot 1 = z = H_1^{(n)}(z,t)$,
$K^2_n \cdot 1 = z^2 = H_2^{(n)}(z,t)$, ..., $K^{n-1}_n \cdot 1 =
z^{n-1} = H_{n-1}^{(n)}(z,t)$, $K^n_n \cdot 1 = z^n+ n!\, t =
H_n^{(n)}(z,t)$, ...,$ K_n^m \cdot 1= H_m^{(n)}(z,t)$, where the
generalized Kampe de Feriet polynomials \cite{Dattoli-01} are\be
H_m^{(n)}(z,t)= m! \sum_{k=0}^{[m/n]}\frac{t^k z^{m-nk}}{k! (m- n
k)!}\ee satisfy the holomorphic Schr\"odinger hierarchy
(\ref{holSchrodingerHierarch})\be \frac{\partial}{\partial t}
H_m^{(n)}(z,t)=\partial_z^n H_m^{(n)}(z,t). \ee The generating
function is given by \be
\sum_{m=0}^{\infty}\frac{k^m}{m!}H_m^{(n)}(z,t)=e^{kz+k^n t}.\ee
From operator representation  \be H_n^{(N)}(z,t)=\exp \left(t
\frac{\partial^N}{\partial z^N} \right)z^n \Rightarrow \Phi(z,t)=
\exp \left( t \frac{\partial^N}{\partial z^N} \right) \psi(z) \ee
we have solution of (\ref{holSchrodingerHierarch})  in terms of
arbitrary analytic function $\psi$. Polynomials $H_m^{(N)}(z,t)$
are connected with the generalized Hermite polynomials
\cite{Srivastava} \be
H_m^{(N)}(z,t)=t^{[m/N]}H_m^{(N)}\left(\frac{z}{\sqrt[N]{t}}\right)
.\ee Then the $k$-th zero $w_n^{(N)k}$ of generalized Hermite
polynomial $H_n^{(N)}$ determine evolution of the corresponding
vortex \be H_n^{(N)}(w_n^{(N)k})=0 \Rightarrow
z_k(t)=w_n^{(N)k}\sqrt[N]{t}.\ee The zeroes are located on the
circle in the plane with time dependent radius. When $t
\rightarrow -t$ position of the vortex rotate on angle $z_k
\rightarrow z_k e^{i\pi/N}$. The Galilean boost hierarchy
(\ref{n-boost}) provides the B\"acklund transformation for n-th
member of anti-holomorphic Burgers hierarchy (\ref{n-Burgers})\be
v=u+\partial_z \ln [z+Nt(\partial_z+u)^{N-1}\cdot 1]. \ee

\section{The Negative Burgers' Hierarchy}

The holomorphic Schr\"odinger hierarchy and corresponding Burgers
hierarchy can be analytically extended to negative values of $N$.
Introducing negative derivative (pseudo-differential) operator
$\partial^{-1}_z$, so that, $
\partial^{-m}_z z^n = \frac{z^{n+m}}{(n+1)...(n+m)}$,
we have the hierarchy \be \partial_{t_{-n}} \Phi =
\partial^{-n}_z \Phi\label{negativeSH}\ee
or differentiating $n$ times, in pure differential form $
\partial_{t_{-n}} \partial^{n}_z\Phi =
 \Phi$.
In terms of $A$ defined by (\ref{compgaugefield}) we have the
negative Burgers hierarchy \be \partial_{t_{-n}} A =
\partial_{\bar z}\frac{1 -
\partial_{t_{-n}}A^n}{A^n}.\ee
For n = 1 we have equation $\partial_{t_{-1}} \Phi =
\partial^{-1}_z \Phi$
or the Helmholz equation $ \partial_{t_{-1}}
\partial_z\Phi =
 \Phi$.
Analytical continuation of the generalized Kampe de Feriet
polynomials to $n = -1$ \cite{Dattoli-01} is given by \be
H_M^{(-1)}(z,t)= M! \sum_{k=0}^{\infty}\frac{t^k z^{M + k}}{k! (M
+ k)!}.\ee Then \be H_M^{(-1)} (z,t) = e^{t
\partial^{-1}_z}H_M^{(-1)} (z,0) \ee \be H_M^{(-1)} (z,0) = z^M
.\ee Moreover higher order functions are generated by the
"negative Galilean boost" \be H_M^{(-1)} (z,t) = (z -
t\partial^{-2}_z)^M H_0^{(-1)} (z,t). \ee Functions $H_M^{(-1)}
(z,t)$ are related with Bessel functions \cite{Dattoli-01}. First,
they are directly related with the Tricomi functions \be C_M (z t)
= \frac{z^{-M}}{M!} H_M^{(-1)} (z,t)\ee determined by the
generating function \be \sum^\infty_{M =-\infty} \lambda^M C_M (x)
= e^{\lambda + x/\lambda}.\ee The last one is connected with
Bessel functions according to \be J_M (x) = \left(
\frac{x}{2}\right)^M C_M (- \frac{x^2}{4}).\ee Then we have
explicitly \be H_M^{(-1)} (z,t)= M! \left(
\frac{-z}{t}\right)^{M/2}  J_M (2 \sqrt{- z t})\ee This provides
solution of the negative (-1) flow Burgers equation \be
\partial_{t} A = \partial_{\bar z}\frac{1 -
\partial_{t}A}{A}\ee
in the form \be A = \frac{(H_M^{(-1)} (\bar z,t))_{\bar
z}}{H_M^{(-1)} (\bar z,t)} = \frac{M}{2\bar z} +
\sqrt{\frac{t}{-\bar z}}\frac{J'_M}{J_M}= \sqrt{\frac{t}{-\bar
z}}\frac{J_{M-1}(2 \sqrt{- \bar z t})}{J_M (2 \sqrt{- \bar z
t})}.\ee For arbitrary member of the negative hierarchy we have
\be H_M^{(-N)} (z,t) = e^{t \partial^{-N}_z}H_M^{(-N)} (z,0) \ee
\be H_M^{(-N)} (z,0) = z^M \ee and relation \be W^{(N)}_M (z
t^{1/N}) = \frac{z^{-M}}{M!} H_M^{(-N)} (z,t)\ee where the
Wright-Bessel functions \cite{Dattoli-01} $W^{(N)}_M (x)$ are
given by generating function \be \sum^\infty_{M =-\infty}
\lambda^M W_M^{(N)} (x) = e^{\lambda + \frac{x}{\lambda^N}}.\ee

\section{Space-Time Noncommutativity}
Now we consider influence of the space-time non-commutativity on
the vortex dynamics. Remarkable is that for two vortices the
problem can be solved explicitly. Non-commutative Burgers'
equation, its linearization by Cole-Hopf transformation and
two-shock soliton collision has been considered in \cite{MP}. Here
we consider the holomorphic heat equation \be
\partial_t \Phi=\nu \,\partial^2_z \Phi \ee
with two-vortex solution in the form  \be
\Phi(z,t)=(z-z_1(t))*(z-z_2(t))\ee where the Moyal product is
defined as
\begin{eqnarray}
 f(t,z)*g(t,z)&=&e^{
i\theta (\partial_t \partial_{z'} - \partial_{t'} \partial_z)} f(t,z) g(t',z')|_{z=z', t=t'}\\
&=& \sum_{n=0}^{\infty}\frac{(i\theta)^n}{n!}\sum_{k=0}^n(-1)^k\left(%
\begin{array}{c}
  n \\
  k \\
\end{array}%
\right) (\partial_t^{n-k}\partial_z^k
f)(\partial_t^{k}\partial_z^{n-k}g)
\end{eqnarray}
Then we have $\theta$ deformed vortex equations
\begin{eqnarray}
\dot{z}_1&=& \frac{-2\nu}{z_1-z_2}-
i \theta\frac{{\ddot z }_1-{\ddot z}_2}{z_1-z_2} \nonumber \\
\dot{z}_2&=& \frac{2\nu}{z_1-z_2}+ i\theta\frac{{\ddot z}_1-{\ddot
z}_2}{z_1-z_2}.
\end{eqnarray}
Adding we have the first integral of motion  $z_1+z_2=C$ - the
center of mass. Choosing beginning of coordinates in the center of
mass we have $C = 0$ and  $z_2=-z_1$. Integrating reduced equation
for $z_1$ and  substituting
 $ z_1=2i\theta Y(t)$ we obtain the  Ricatti equation\be \dot Y+Y^2
=\frac{\nu}{2 \theta^2}(t-t_0) .\ee This can be linearized by $Y =
\dot \psi/\psi$ in terms of the Airy Equation \be \ddot \psi =
\frac{\nu}{2\theta^2}(t-t_0)\,\psi .\ee The solution is \be
z_1(t)=2i\theta \beta \frac{Ai'(\beta (t - t_0))}{Ai(\beta (t -
t_0))} = -i\sqrt{2\nu
(t-t_0)}\frac{K_{2/3}(\frac{\sqrt{2\nu}}{3\theta}(t-t_0)^{3/2})}
{K_{1/3}(\frac{\sqrt{2\nu}}{3\theta}(t-t_0)^{3/2})} \ee where
$\beta = (\nu/2\theta^2)^{1/3}$ and $K_n$ are modified Bessel
functions of fractional order. This solution should be compared
with the undeformed one (\ref{2vortex}). The noncommutative
corrections are coming from the ratio of two Bessel functions
depending on $\theta$. Using asymptotic form of Airy function we
have correction in the form
\begin{eqnarray}
  z_1(t) = - z_2 (t)&
\approx & - i\sqrt{2 \nu (t-t_0)}-\frac{i\theta}{2(t-t_0)}
\end{eqnarray}
as $t \rightarrow +\infty$. As easy to see the correction is
independent of diffusion coefficient and has the global character.
\section{Acknowledgements}
This work was partially supported by TUBITAK under the Grant No.
106T447.


\begin{thebibliography}{999}
\parindent=.6em



\bibitem{MPS1}
L. Martina, O. K. Pashaev, G. Soliani, J. Phys. A: Math. Gen., 27,
(1994).

\bibitem{Mermin}N. D. Mermin, T. Ho, Phys Rev Lett, 36, No.11,(1976); Phys Rev E,
Vol.21, No.11,(1980).

\bibitem{MPS3}L. Martina, O. K. Pashaev, G. Soliani, Theor.
Math. Phys. 99, 726-732, (1994).

\bibitem{Ishimori}Y. Ishimori, Prog. of Theor. Phys., Vol.72, No.1, July (1984).

\bibitem{Lav} M. A. Lavrentiev, B. V. Shabat, Problems of
Hydrodynamics and Their Mathematical Models. Nauka, Moscow,
(1973)(in Russian).

\bibitem{Chood}D. V. Choodnovsky, G. V. Choodnovsky, Il Nuovo Cimento, Vol. 40B, No. 2, 11 August (1977).


\bibitem{Calogero}F. Calogero, Il Nuovo Cimento, Vol.43B, No.2, 11 February (1978).

\bibitem{CalSanSom}F. Calogero, D. Gomez- Ullate, P. M. Santini,
M. Sommacal, J. Phy. A: Math. Gen. 38, (2005), 8873- 8896.


\bibitem{Perelomov}A. M. Perelomov, Integrable Systems of Classical Mechanics and Lie
Algebras, Vol. 1, Birkh\"auser Verlag, (1990).


\bibitem{Calogerob} F. Calogero, Classical Many-Body Problems
Amenable to Exact Treatments, Springer, (2001).

\bibitem{Wiegmann} A. B. Abanov, P. B. Wiegmann, Phys. Rev. Lett.
95 (2005) 076402

\bibitem{Bonami} A. Bonami, F. Bouchut, E. Cepa, D. Lepingle, J.
Functional Analysis, 165 (1999) 390-406

\bibitem{Dattoli-97}G. Dattoli, P. L. Ottaviani, A. Torre and L. Vazquez, Revista del Nuovo
Cimenti, V.20, (1997) 1-128.

\bibitem{Srivastava} H. M. Srivastava, Nederl. Akad. Wetensch.
Proc. Ser. A 79, (1976) 457-461

\bibitem{Dattoli-01}G. Dattoli,
P. E. Ricci,C. Cesarano,  Applicable Analysis, Vol. 80, (2001).
379-384.






\bibitem{MP}L. Martina, O. K. Pashaev, in Nonlinear Physics: Theory
and Experiment. II, World Scientific, (2003), 83-88.


\end{thebibliography}
\end{document}